\documentclass[11pt]{article}

\usepackage[utf8]{inputenc}
\usepackage[T1]{fontenc}
\usepackage[a4paper,margin=1in]{geometry}
\usepackage{amsmath,amssymb}
\usepackage{graphicx}
\usepackage{booktabs,longtable,array,multirow}
\usepackage{calc}
\usepackage{xcolor}
\usepackage{hyperref}
\usepackage{url}

\hypersetup{hidelinks}

\title{Automatic Adjustment of HPA Parameters and Attack Prevention in Kubernetes Using Random Forests}
\author{Huah Yong Chan\\
School of Computer Sciences, Universiti Sains Malaysia\\
\texttt{hychan@usm.my}
\and
Hanlin Zhou\\
School of Computer Sciences, Universiti Sains Malaysia\\
Xiamen Institute of Software Technology, China\\
\texttt{zhouhanlin1@student.usm.my}
\and
Jingfei Ni\\
Information Management Department of Manzhouli Customs, China\\
\texttt{morgan\_ni@163.com}
\and
Mengchun Wu\\
College of Information Science and Engineering, Jimei University, China\\
\texttt{wumengchun04@gmail.com}
\and
Qing Deng\\
School of Computer Sciences, Universiti Sains Malaysia\\
\texttt{dengqing@student.usm.my}}
\date{}

\begin{document}
\maketitle

\begin{abstract}
In this paper, HTTP status codes are used as custom metrics within the
HPA as the experimental scenario. By integrating the Random Forest
classification algorithm from machine learning, attacks are assessed and
predicted, dynamically adjusting the maximum pod parameter in the HPA to
manage attack traffic. This approach enables the adjustment of HPA
parameters using machine learning scripts in targeted attack scenarios
while effectively managing attack traffic. All access from attacking IPs
is redirected to honeypot pods, achieving a lower incidence of 5XX
status codes through HPA pod adjustments under high load conditions.
This method also ensures effective isolation of attack traffic,
preventing excessive HPA expansion due to attacks. Additionally,
experiments conducted under various conditions demonstrate the
importance of setting appropriate thresholds for HPA adjustments.
\end{abstract}

\noindent\textbf{Keywords:} Kubernetes, HPA, Security, Random Forest

\vspace{1em}
\hypertarget{introduction}{%
\section{Introduction}\label{introduction}}

In the past decade, cloud computing technology has developed
rapidly[1], and container-based Docker services have
become increasingly common, particularly in providing web services.
Using Nginx containers allows for the quick setup of web
services[2]. In the realm of container management,
Kubernetes is the most commonly used platform for container management
and orchestration, efficiently handling container management at the pod
level. Among the many features in Kubernetes, the auto-scaling function
is widely used. The Horizontal Pod Autoscaler (HPA)[3]
can quickly scale pods up and down. For Nginx pods, the relevant metric
is the user's request volume. However, distinguishing between normal
user requests and abnormal attack traffic is challenging. Abnormal
requests can cause the HPA to scale the number of containers to its
maximum, thereby consuming cloud platform resources[4].

To address this, HPA can be used for automatic pod scaling, but judge
whether requests are abnormal remains a challenge[5].
This paper proposes using machine learning to predict attack behaviors
and dynamically adjust the HPA parameters accordingly.

\hypertarget{custom-metrics-hpa}{%
\subsection{Custom Metrics HPA}\label{custom-metrics-hpa}}

There are three types of resource scaling in Kubernetes: HPA, VPA
(Vertical Pod Autoscaler), and CA (Cluster Autoscaler). By default, HPA
adjusts the number of pods based on resource usage such as CPU and
memory[6]. However, sometimes CPU or memory usage may not
accurately reflect the service status of the corresponding pod. For
instance, in Nginx containers aimed at responding to user HTTP
requests[7], if there is a service anomaly due to reasons
like program crashes or upstream service issues, the CPU and memory
usage of the container itself or its upstream services may still appear
normal[8], but users may receive a large number of 5xx
status codes. This leads to two issues when using HPA with default
metrics. First, HPA cannot scale itself based on the values of the pod.
Second, HPA only has CPU and memory as default metrics, and using custom
metrics requires the installation of an additional API server.

\hypertarget{dynamic-adjustment-of-hpa-parameters}{%
\subsection{Dynamic Adjustment of HPA
Parameters}\label{dynamic-adjustment-of-hpa-parameters}}

Moreover, addressing the increasingly severe issue of cyberattacks is
essential, specifically preventing resource wastage caused by the HPA's
automatic pod expansion in response to attacks. It is crucial to
distinguish between normal request traffic and attack traffic. For
instance, a DDoS attack can generate excessive traffic, exceeding the
threshold and triggering the HPA, thereby allowing a short-term DDoS
attack to have a prolonged impact[9].

This paper takes targeted scanning attacks as an example and
designs a program capable of identifying whether requests are normal
traffic or scanning directory attack traffic. If an attack is detected,
it is redirected to honeypot pods that do not scale[10],
while other traffic is normally served by Nginx pods.

Under typical usage, HPA parameters are fixed after creation. However,
this paper explores the use of dynamic HPA parameters, setting different
HPA parameters at various stages of an attack. For instance, when there
is no attack, the maximum number of Nginx pods is set to 5. During an
attack, the maximum pod value is initially reduced to 1 to lower the
scaling cost caused by the attack. In a Kubernetes environment, many
parameters can be adjusted through APIs. This paper attempts to use
machine learning to determine the presence of an attack and whether it
is necessary to adjust HPA configuration parameters.

\hypertarget{random-forest-classification}{%
\subsection{Random Forest
Classification}\label{random-forest-classification}}

Random Forest is one of the most commonly used machine learning
algorithms[11]. It includes decision tree classifiers,
where the algorithm generates various trees based on the values of
sample cases, forming a forest. The final training result is then
integrated from these trees in the forest. Using Random Forest for data
with a high degree of randomness has certain advantages. In this paper,
Random Forest is used solely for classifying directory
attacks[12]. Machine learning is employed to determine
the HPA parameters, specifically identifying whether certain requests
are target scanning attacks and predicting the likelihood of future
attacks.

This paper presents an experimental attempt to utilize cloud computing
and machine learning, with its primary innovation lying in the use of
machine learning to detect attacks. This detection is integrated into an
HPA based on HTTP status codes as custom metrics, enabling dynamic
adjustment of the HPA's maximum pod parameter based on attack
assessments. Additionally, IPs with abnormal traffic are redirected to a
honeypot isolation space outside the control of the HPA. This approach
achieves automated and customized container management, preventing
resource waste or service unavailability caused by targeted scanning
attacks.

\hypertarget{methodology}{%
\section{METHODOLOGY}\label{methodology}}

This study aims to explore the application of machine learning in the
automatic scaling of cloud computing containers. It employs a custom
architecture to implement HPA using HTTP status codes as custom metrics.
Additionally, Random Forest from machine learning is utilized to
identify and assess attacks, achieving more advanced traffic
classification for pod management.

\hypertarget{implementation-of-hpa-using-http-status-codes-as-custom-metrics}{%
\subsection{Implementation of HPA Using HTTP Status Codes as Custom
Metrics}\label{implementation-of-hpa-using-http-status-codes-as-custom-metrics}}

Since the default HPA functionality in Kubernetes only supports using
CPU and memory as default metrics, this section describes the
implementation of an HPA using HTTP status codes as custom metrics. This
is achieved through a two-tier architecture that exports HTTP status
codes and queries Prometheus[13]. The growth rate of 5xx
status codes over a 5-minute period is used as the HPA trigger threshold
like SLA[14].

To implement this, a dual-layer Nginx pod architecture is used. The
first layer of Nginx pods acts as a reverse proxy server, providing
external network service ports. This layer forwards requests to the
second layer of Nginx servers. The first layer also records HTTP status
codes and allows Prometheus and custom metrics APIs to access the
recorded status code data[15]. The second layer of Nginx
pods serves as the upstream server for the first layer, providing actual
services and being the target for HPA scaling since it is the actual
service pod.

Specifically, the first layer Nginx pod uses the nginx-module-vts (Nginx
Virtual Host Traffic Status module, VTS) to log user request data. This
module can convert the data format to JSON. Additionally, the
nginx-vts-exporter module is used to extract VTS statistics from Nginx
and export them in a format readable by Prometheus.

The second layer Nginx pod is a standard Nginx Docker container that
only provides homepage services. In this study, since the focus is on
the load of Nginx pods and using user status codes as custom metrics for
HPA, the upstream service server uses Nginx. Refer to Figure 1 for
details.

\includegraphics[width=5.60208in,height=2.40069in]{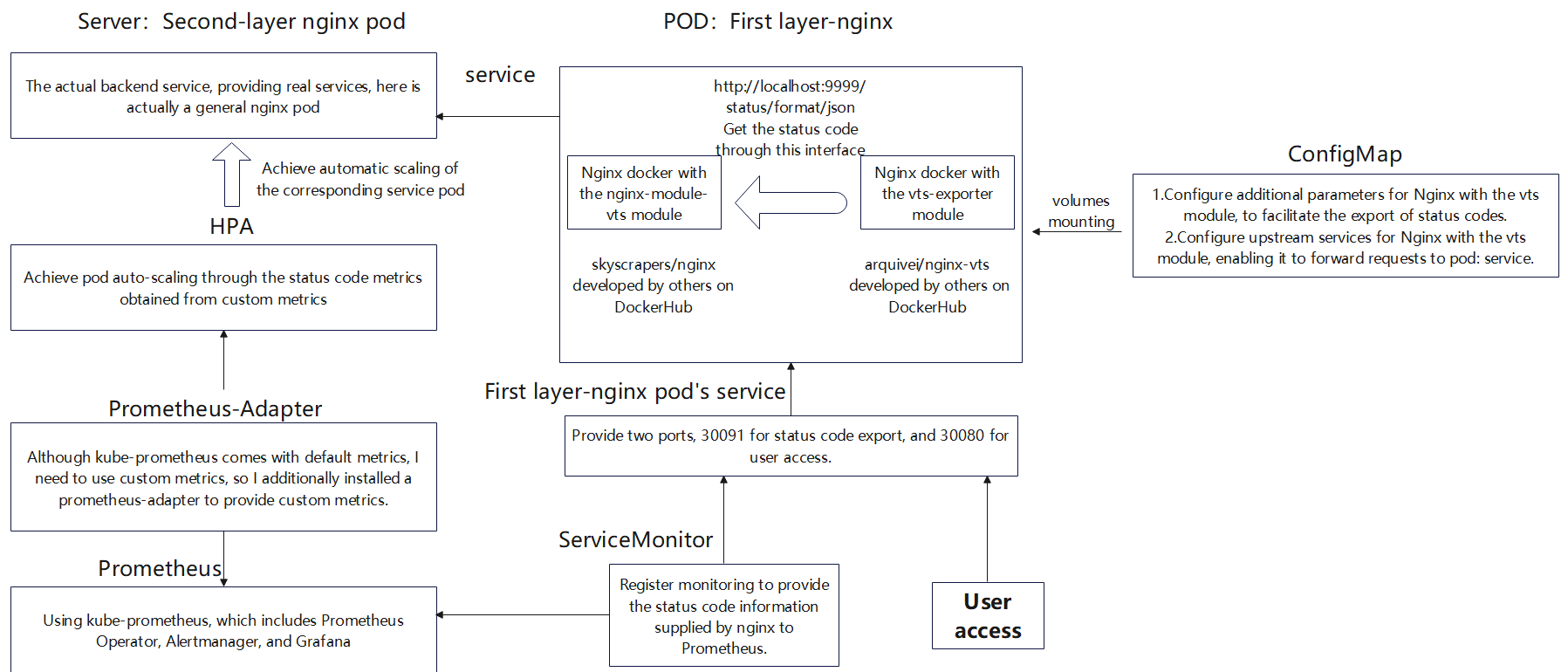}

Figure 1: HPA Architecture Using HTTP Status Codes as Custom Metrics

These services will be implemented through Kubernetes services. The
first layer pods will open two ports: one port for outputting status
codes and another port as the user access interface for the Nginx
service. This setup allows the first layer pods to expose the recorded
Nginx status codes through a port, including various data such as 5xx
status codes and 4xx status codes, among others.

\hypertarget{using-random-forest-to-identify-attack-traffic}{%
\subsection{Using Random Forest to Identify Attack
Traffic}\label{using-random-forest-to-identify-attack-traffic}}

In this section, for the sake of experimentation, a Python script will
be used to run a Random Forest classifier to determine whether certain
requests are attacks or normal access. The script will also identify the
top 10 IPs associated with directory scanning attacks within a 5-minute
window and predict their future attack probability.

Since this is a test scenario, access is also simulated using a Python
script. Due to the virtual machine environment, the generated virtual
IPs are stored in the X-Forwarded-For header. In the logs, this IP
appears at the end, so the detection model will use the X-Forwarded-For
IP as the source IP to determine if there is a likelihood of attack from
that IP[16].

First, a scheduled task exports the generated logs to a specific
file[17], which serves as the data-set for model
training. The log data includes entries with connection refusals due to
timeouts or access anomalies, necessitating data preprocessing. Logs are
read according to their format, and improperly formatted logs are
filtered out. All log entries are then converted into a
DataFrame[18].

Next, data labeling is performed. Access status codes 404 and 403 are
marked as abnormal access, and error logs are labeled as 1, while other
logs are labeled as 0. The access logs and error logs are then merged.
Using the IP as the action feature for classification, 80\% of the data
is used as the training set and 20\% as the test set. The Random Forest
classifier model is trained on this data and saved. The model uses 100
trees and sets the random seed to 42. After training, the model is
evaluated using the test set, and the training results are outputted.
The overall process is shown in Figure 2.

\includegraphics[width=6.07986in,height=0.22847in]{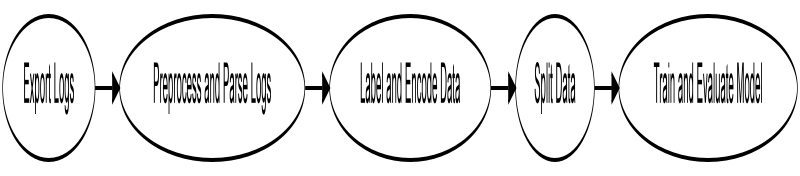}

Figure 2: Random Forest Model Steps

In this study, attack scripts will utilize the Faker library to generate
requests. The attacking IPs can be set as either fixed or random
IPs[19]. For experimental validation, fixed IPs are used
to perform scans on common targets. A total of 200 IPs are used, with
only some IPs involved in the attacks. Not all requests are attacks;
only a subset of accesses will scan for sensitive targets such as admin
and login.

The following Figure 3 shows a screenshot of the experiment running a
Random Forest classifier using a Python script. It displays the training
results and the top 10 attacking IPs along with their predicted future
attack probabilities.

\includegraphics[width=2.25694in,height=2.425in]{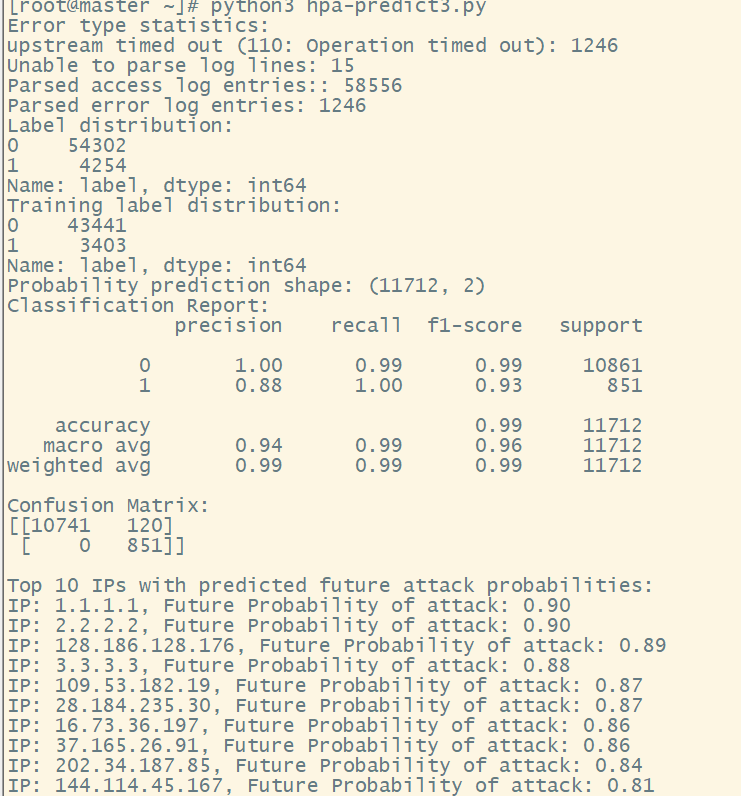}

Figure 3: Example of the result of executing a random forest script

\hypertarget{implementation-of-dynamic-adjustment-of-hpa-parameters}{%
\subsection{Implementation of Dynamic Adjustment of HPA
Parameters}\label{implementation-of-dynamic-adjustment-of-hpa-parameters}}

In normal access scenarios, the HPA requires a minimum number of pods
(minReplicas) and a maxReplicas. When the trigger threshold is not met,
the minimum number of pods is generated. For instance, in this study,
the HPA is set to trigger when the number of 5xx status codes increases
by more than 50 within five minutes, promptly adjusting the
corresponding pods to the maximum value[20].

However, some requests may be attack traffic. In this case, the values
of maxReplicas or minReplicas can be dynamically adjusted. This
adjustment can be achieved by calling the API interface. By using a
Python script to load the kube configuration, the HPA object's
information can be read and the HPA parameters can be updated. Two
Python scripts were created: one to adjust maxReplicas to 1 and the
other to adjust it to 5. This allows the previous machine learning
Python code to dynamically modify the HPA parameters based on training
results.

For example, when an attack volume greater than 1\% is detected, the
maxReplicas value is reduced to prevent resource wastage due to HPA
scaling caused by the attack. This can be integrated into the previous
Python code for one-click management.

\hypertarget{management-of-attack-traffic}{%
\subsection{Management of Attack
Traffic}\label{management-of-attack-traffic}}

In the context of using HPA with dynamic parameter adjustment,
continuously setting a low maximum pod value due to an attack can impact
other users' access. This paper introduces an additional layer of Nginx
pods to manage traffic redirection, directing attack traffic to specific
isolated honeypot pods. Meanwhile, the HPA will scale based on normal
user behavior, and attack traffic will be redirected to another Nginx
pod. This approach separates normal request traffic from attack traffic,
allowing the HPA to operate without being affected by attack traffic. It
also unifies the management of attack traffic and false positives,
limiting their resource usage and facilitating better analysis of their
characteristics through log analysis.

The following Figure 4 illustrates the overall architecture after
incorporating attack traffic management. User traffic enters the Nginx
pod providing external services through the Third layer-Nginx pod. By
default, all traffic passes through the service to the First layer-Nginx
pod, which then forwards the traffic to the required service layer, the
Second-layer Nginx pod. If an attack is detected within the current user
requests, a machine learning script is used for attack detection. The
script dynamically updates the Third layer-Nginx pod's configmap
configuration file with the attacking IP, writing the corresponding
X-Forwarded-For IP into the Nginx configuration file. As a result,
attack traffic is directed to the isolated honeypot pods, while other
pods handle normal request traffic. HPA parameters are adjusted
accordingly.

\includegraphics[width=6.07431in,height=2.55972in]{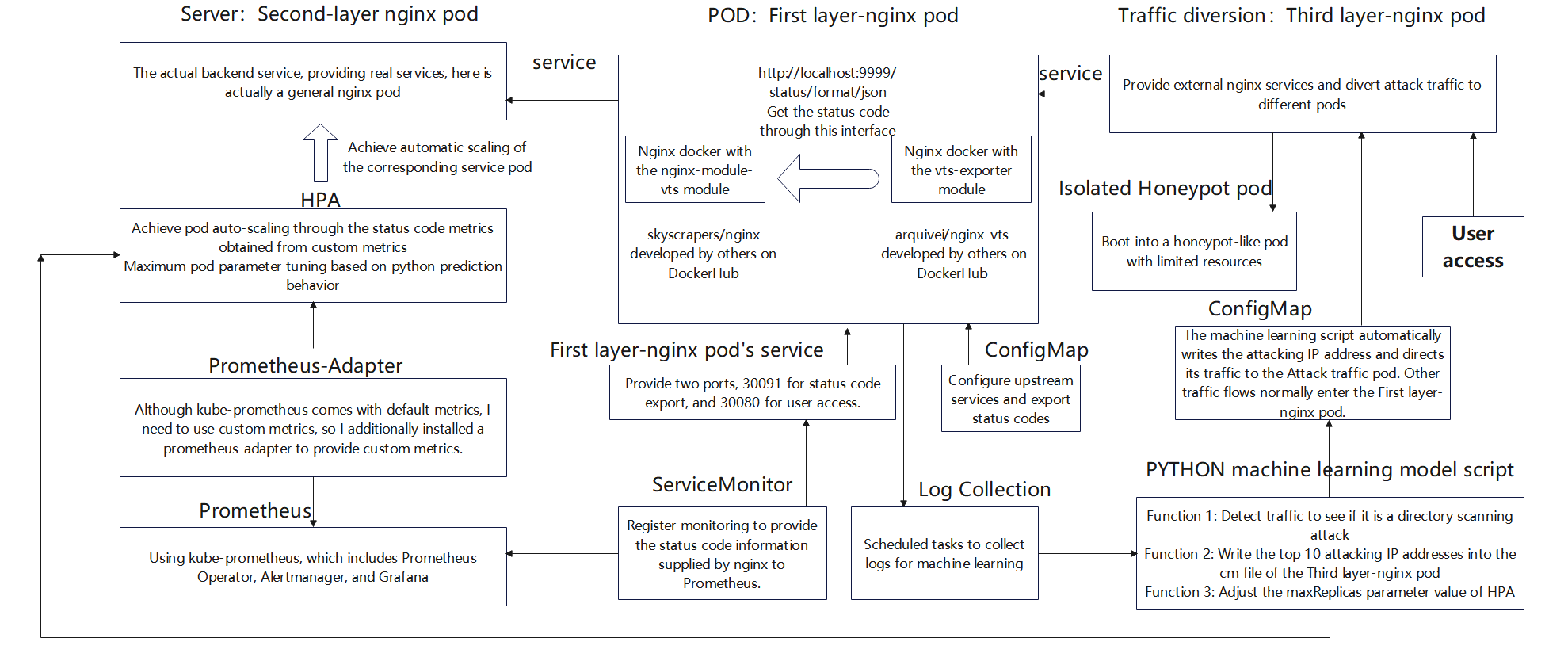}

Figure 4: Dynamically controlled attack prevention architecture of
machine learning and HPA

\hypertarget{performance-analysis}{%
\section{PERFORMANCE ANALYSIS}\label{performance-analysis}}

In the research chapters of this article, the focus will be on the study
of dynamically adjusting HPA parameters based on machine learning. The
experimental scenario involves attack scanning on target web
applications and managing the traffic of attacking IPs. This chapter
will compare user access performance and infrastructure cost management
under different conditions when executing machine learning scripts,
analyzing the impact of Python-based machine learning scripts on the
overall HPA scenario.

Since this paper involves exploratory experiments, both normal requests
and attack requests are simulated using Python. The concurrency level is
set to 200 due to the limitations of the experimental platform
resources. When the CPU resources for the Nginx pod are set to 30m,
concurrency exceeding 200 triggers the HPA (Horizontal Pod Autoscaler)
for scaling, as requests surpass the maximum service capacity of a
single pod. Additionally, 200 concurrent requests represent the service
capacity that 5 pods can handle. Although the concurrent request count
is 200, the actual request rate is 400-600 requests per second. This
discrepancy occurs because the requests are for the Nginx default
homepage, which can be responded to within a second, thus resulting in
more than 200 requests per second. The total request count of 200,000
was chosen to simplify the experimental steps while ensuring
reliability.

In the designed experiment, 200 IP addresses are used to access the
target URL, with 10 of these IPs potentially exhibiting attack behavior.
These attack behaviors include probing for common targets directories
like admin, data, and login. The remaining 190 IPs perform normal
accesses, only requesting the homepage. Two attack probabilities are
considered: 12\% and 3\%. For the 12\% attack probability scenario, 80\%
of normal traffic is generated by 190 random IPs, and 20\% of traffic is
from 10 random IPs, with 60\% of this traffic being directory scanning
attacks and 40\% normal access (20\% * 60\% = 12\%). For the 3\% attack
probability scenario, 20\% of the potentially attacking traffic from 10
IPs has only 15\% attack requests (20\% * 15\% = 3\%), making it
meaningful to predict future attacks as not all requests are attacks.

The choice of 12\% and 3\% attack probabilities is to explore the
dynamic combination of machine learning and HPA for attack isolation.
The HPA maxReplicas trigger values are derived from exploratory
validation, covering boundary conditions for triggering and not
triggering.

This exploratory experiment employs a Python machine learning script
using a random forest to determine if an IP's request is an attack and
to predict future attack probabilities. This module can be replaced with
other machine learning models, and different types of attacks can be
simulated. Here, a random forest is used to determine if a request is a
directory scanning attack. The architecture can be generalized and
applied to other scenarios by replacing the Python machine learning
script with different models to identify and isolate various attacks.
This paper demonstrates the feasibility of the framework using directory
scanning as an example, highlighting its applicability across different
contexts.

The experiment employs targeted attacks as the attack method. Below
Table 1 is the functions and specific data for such requests and
responses. The Python machine learning script executes at the 3rd and
5th minutes following the simulated attack. Each time, the attacking IP
is changed. The HPA target pod CPU limit is set to 30m to trigger HPA
based on 5XX status codes as custom metrics under high concurrency
conditions[21].

Table 1: Experimental variable conditions

\begin{longtable}[]{@{}
  >{\raggedright\arraybackslash}p{(\columnwidth - 8\tabcolsep) * \real{0.1228}}
  >{\raggedright\arraybackslash}p{(\columnwidth - 8\tabcolsep) * \real{0.0981}}
  >{\raggedright\arraybackslash}p{(\columnwidth - 8\tabcolsep) * \real{0.3723}}
  >{\raggedright\arraybackslash}p{(\columnwidth - 8\tabcolsep) * \real{0.2150}}
  >{\raggedright\arraybackslash}p{(\columnwidth - 8\tabcolsep) * \real{0.1918}}@{}}
\toprule
\begin{minipage}[b]{\linewidth}\raggedright
Condition
\end{minipage} & \begin{minipage}[b]{\linewidth}\raggedright
Attack Probability
\end{minipage} & \begin{minipage}[b]{\linewidth}\raggedright
HPA MaxReplicas Trigger Value
\end{minipage} & \begin{minipage}[b]{\linewidth}\raggedright
HPA Trigger Conditions
\end{minipage} & \begin{minipage}[b]{\linewidth}\raggedright
Total Requests and Concurrent Requests
\end{minipage} \\
\midrule
\endhead
1 & 12\% & The attack rate in the last 5 minutes is 10\%. &
\multirow{6}{*}{50 5XX status codes have been added in the last 5
minutes} & \multirow{6}{*}{200000, 200} \\
2 & 12\% & The attack rate in the last 5 minutes is 20\%. \\
3 & 3\% & The attack rate in the last 5 minutes is 1\%. \\
4 & 3\% & The attack rate in the last 5 minutes is 5\%. \\
5 & 3\% & The attack rate in the last 1 minute is 1\%. \\
6 & 3\% & The attack rate in the last 1 minute is 5\%. \\
\bottomrule
\end{longtable}

The experimental data is presented as follows. Figures 5 through 10 are
Grafana display charts that show the status code data of the second
layer of pods, which are the actual service pods. To ensure data
reliability, each experiment condition is repeated three times, and the
average value is taken. The displayed charts represent the results of
the first run for each condition.

\includegraphics[width=2.49028in,height=2.27778in]{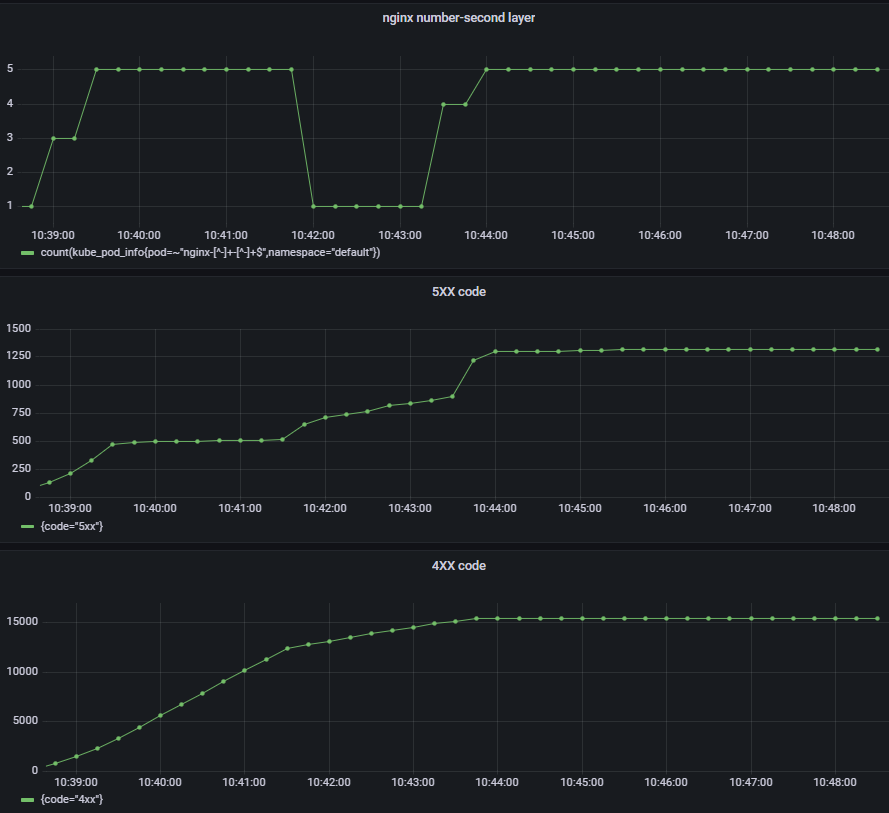}
\includegraphics[width=2.49236in,height=2.27986in]{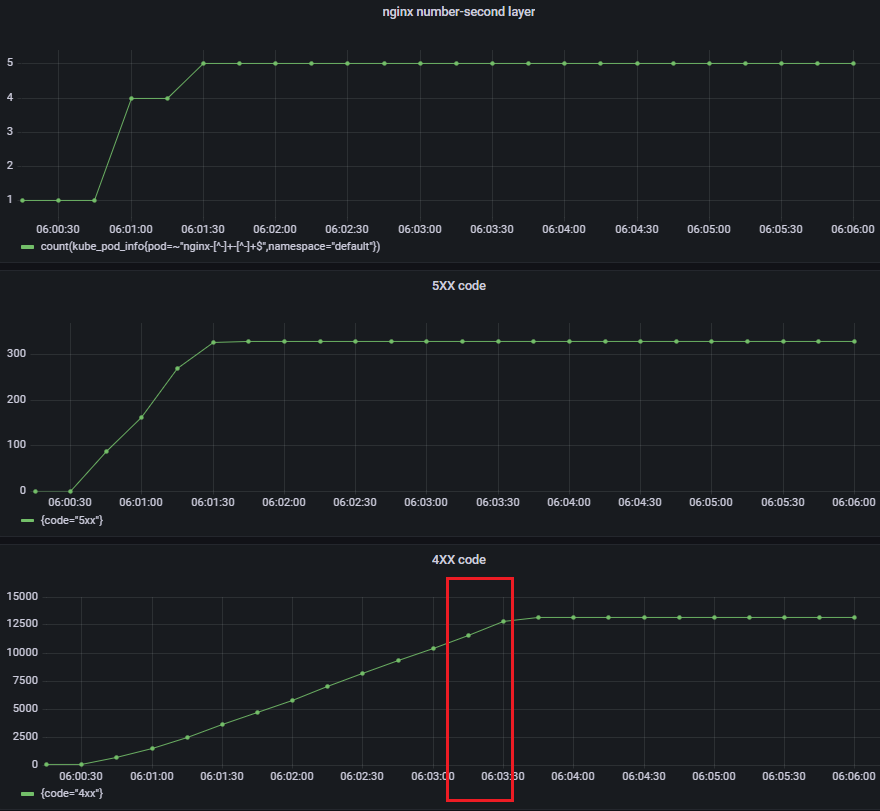}

Figure 5: The results of Condition 1 in Grafana Figure 6: The results of
Condition 2 in Grafana

Table 2: Result of condition 1

\begin{longtable}[]{@{}
  >{\raggedright\arraybackslash}p{(\columnwidth - 14\tabcolsep) * \real{0.0803}}
  >{\raggedright\arraybackslash}p{(\columnwidth - 14\tabcolsep) * \real{0.0921}}
  >{\raggedright\arraybackslash}p{(\columnwidth - 14\tabcolsep) * \real{0.1649}}
  >{\raggedright\arraybackslash}p{(\columnwidth - 14\tabcolsep) * \real{0.1249}}
  >{\raggedright\arraybackslash}p{(\columnwidth - 14\tabcolsep) * \real{0.1735}}
  >{\raggedright\arraybackslash}p{(\columnwidth - 14\tabcolsep) * \real{0.1020}}
  >{\raggedright\arraybackslash}p{(\columnwidth - 14\tabcolsep) * \real{0.1102}}
  >{\raggedright\arraybackslash}p{(\columnwidth - 14\tabcolsep) * \real{0.1522}}@{}}
\toprule
\begin{minipage}[b]{\linewidth}\raggedright
Number
\end{minipage} & \begin{minipage}[b]{\linewidth}\raggedright
Condition
\end{minipage} & \begin{minipage}[b]{\linewidth}\raggedright
Nginx Attacks Received(Second layer pod)
\end{minipage} & \begin{minipage}[b]{\linewidth}\raggedright
5XX Number
\end{minipage} & \begin{minipage}[b]{\linewidth}\raggedright
Honeypot Attacks Received(Isolating pods)
\end{minipage} & \begin{minipage}[b]{\linewidth}\raggedright
Total Request Time
\end{minipage} & \begin{minipage}[b]{\linewidth}\raggedright
The First F1-Score Attack
\end{minipage} & \begin{minipage}[b]{\linewidth}\raggedright
The Average Attack Rate of The First Attacking Ip In The Future
\end{minipage} \\
\midrule
\endhead
1 & 1 & 11367 & 1360 & 12455 & 452 & 0.99 & 0.974 \\
2 & 1 & 10233 & 1500 & 13463 & 444 & 0.99 & 0.975 \\
3 & 1 & 10934 & 1443 & 12643 & 425 & 0.98 & 0.974 \\
Average & & 10844 & 1434 & 12853 & 431 & 0.99 & 0.974 \\
\bottomrule
\end{longtable}

At the beginning of a large number of requests, a rapid increase of pods
is observed, indicating the activation of the auto-scaling function to
handle high concurrent traffic. However, due to the HPA's maxReplicas
trigger value being set to an attack ratio of 10\% over the last five
minutes, the maximum HPA value is set to 1 after the first trigger. As a
result, the business pod is scaled down to 1, leading to a rapid decline
in system service capacity, and the number of 5XX errors continues to
rise. This continues until the second execution of the machine learning
script identifies that the attack ratio has fallen below the
corresponding threshold, restoring the HPA max value to 5. Consequently,
the system service capacity quickly recovers, and the number of 5XX
errors caused by insufficient resources no longer increases.

In the figure, the corresponding 4XX ratio continues to increase after
the execution of the first script. This is due to the high traffic
volume leading to 499 status codes, which are not attack-related
traffic.

Table 3: Result of condition 2

\begin{longtable}[]{@{}
  >{\raggedright\arraybackslash}p{(\columnwidth - 14\tabcolsep) * \real{0.0886}}
  >{\raggedright\arraybackslash}p{(\columnwidth - 14\tabcolsep) * \real{0.0908}}
  >{\raggedright\arraybackslash}p{(\columnwidth - 14\tabcolsep) * \real{0.1579}}
  >{\raggedright\arraybackslash}p{(\columnwidth - 14\tabcolsep) * \real{0.1249}}
  >{\raggedright\arraybackslash}p{(\columnwidth - 14\tabcolsep) * \real{0.1735}}
  >{\raggedright\arraybackslash}p{(\columnwidth - 14\tabcolsep) * \real{0.1020}}
  >{\raggedright\arraybackslash}p{(\columnwidth - 14\tabcolsep) * \real{0.1102}}
  >{\raggedright\arraybackslash}p{(\columnwidth - 14\tabcolsep) * \real{0.1522}}@{}}
\toprule
\begin{minipage}[b]{\linewidth}\raggedright
Number
\end{minipage} & \begin{minipage}[b]{\linewidth}\raggedright
Condition
\end{minipage} & \begin{minipage}[b]{\linewidth}\raggedright
Nginx Attacks Received(Second layer pod)
\end{minipage} & \begin{minipage}[b]{\linewidth}\raggedright
5XX Number
\end{minipage} & \begin{minipage}[b]{\linewidth}\raggedright
Honeypot Attacks Received(Isolating pods)
\end{minipage} & \begin{minipage}[b]{\linewidth}\raggedright
Total Request Time
\end{minipage} & \begin{minipage}[b]{\linewidth}\raggedright
The First F1-Score Attack
\end{minipage} & \begin{minipage}[b]{\linewidth}\raggedright
The Average Attack Rate of The First Attacking Ip In The Future
\end{minipage} \\
\midrule
\endhead
1 & 2 & 12140 & 314 & 11422 & 333 & 0.99 & 0.983 \\
2 & 2 & 11085 & 560 & 11987 & 353 & 0.96 & 0.922 \\
3 & 2 & 11991 & 592 & 11615 & 342 & 0.99 & 0.977 \\
Average & & 11739 & 489 & 11675 & 342 & 0.98 & 0.961 \\
\bottomrule
\end{longtable}

The data analysis indicates that at the onset of a large number of
requests, there is a rapid increase in requests, triggering the HPA
auto-scaling function to handle the high concurrency. However, since the
HPA's maxReplicas trigger value is set to an attack ratio of 20\% over
the last five minutes, after the first execution of the script, the IP
addresses associated with attack traffic are redirected to the
corresponding isolation zone, resulting in no further increase in the
number of 4XX errors.

Moreover, due to the high maxReplicas trigger value of the HPA, neither
the first nor the second execution of the script resulted in a reduction
of the maxReplicas value. Consequently, the overall availability of the
system remains high, with a very low proportion of 5XX errors and a
relatively low total request time.The red part is the script execution,
and 4XX traffic starts to enter the honeypot pod.

\includegraphics[width=2.49236in,height=2.27986in]{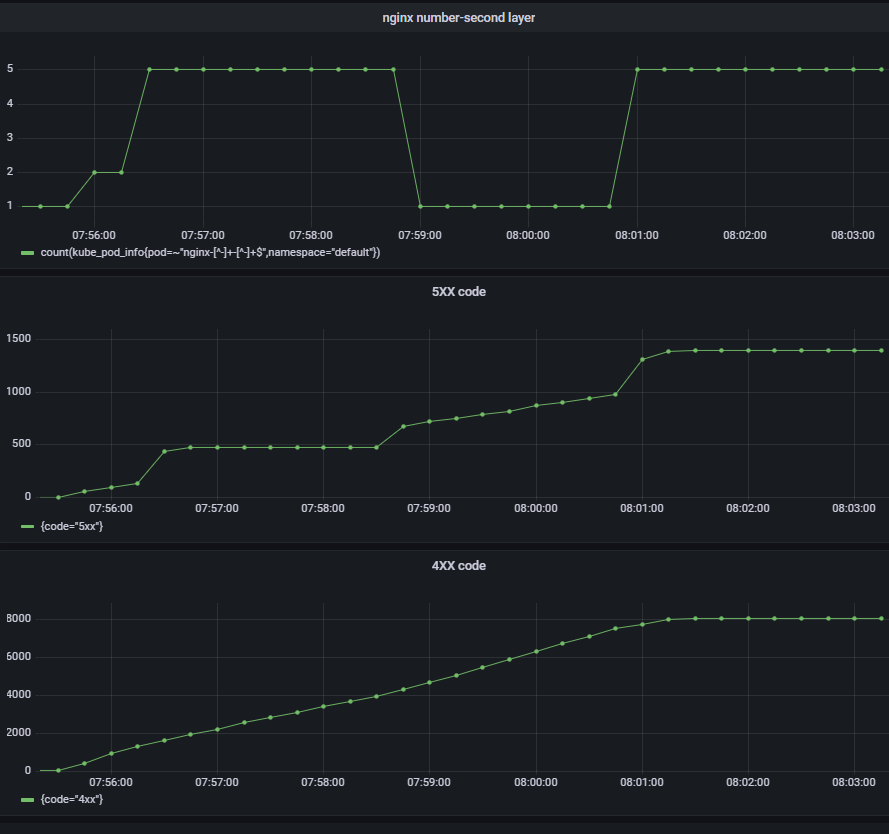}
\includegraphics[width=2.49236in,height=2.27986in]{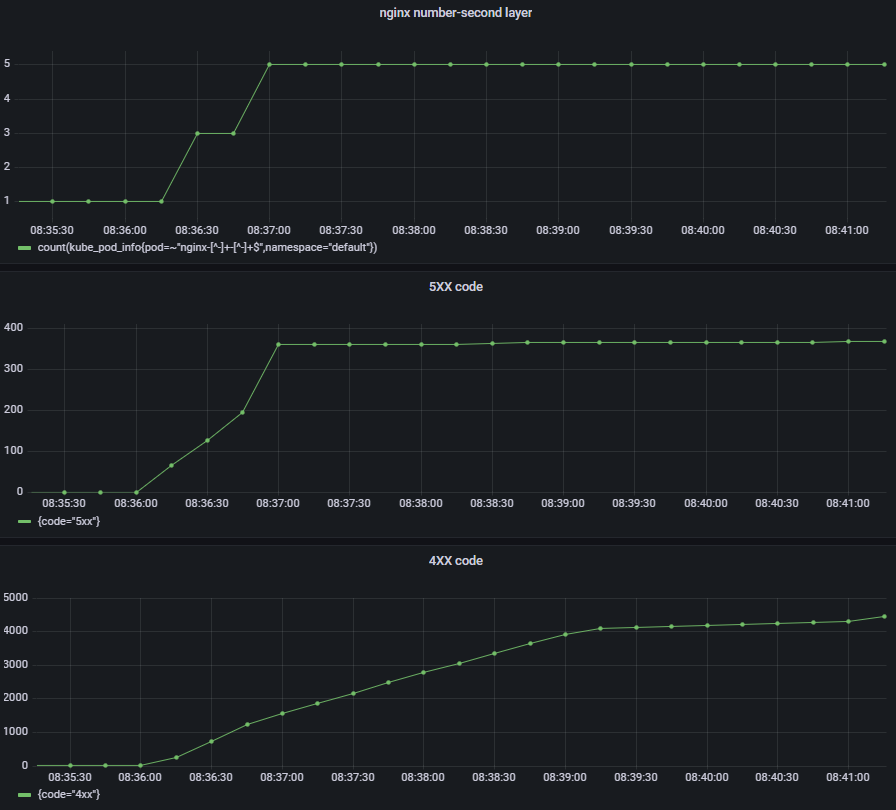}

Figure 7: The results of Condition 3 in Grafana Figure 8: The results of
Condition 4 in Grafana

Table 4: Result of condition 3

\begin{longtable}[]{@{}
  >{\raggedright\arraybackslash}p{(\columnwidth - 14\tabcolsep) * \real{0.0886}}
  >{\raggedright\arraybackslash}p{(\columnwidth - 14\tabcolsep) * \real{0.0908}}
  >{\raggedright\arraybackslash}p{(\columnwidth - 14\tabcolsep) * \real{0.1579}}
  >{\raggedright\arraybackslash}p{(\columnwidth - 14\tabcolsep) * \real{0.1249}}
  >{\raggedright\arraybackslash}p{(\columnwidth - 14\tabcolsep) * \real{0.1735}}
  >{\raggedright\arraybackslash}p{(\columnwidth - 14\tabcolsep) * \real{0.1020}}
  >{\raggedright\arraybackslash}p{(\columnwidth - 14\tabcolsep) * \real{0.1102}}
  >{\raggedright\arraybackslash}p{(\columnwidth - 14\tabcolsep) * \real{0.1522}}@{}}
\toprule
\begin{minipage}[b]{\linewidth}\raggedright
Number
\end{minipage} & \begin{minipage}[b]{\linewidth}\raggedright
Condition
\end{minipage} & \begin{minipage}[b]{\linewidth}\raggedright
Nginx Attacks Received(Second layer pod)
\end{minipage} & \begin{minipage}[b]{\linewidth}\raggedright
5XX Number
\end{minipage} & \begin{minipage}[b]{\linewidth}\raggedright
Honeypot Attacks Received(Isolating pods)
\end{minipage} & \begin{minipage}[b]{\linewidth}\raggedright
Total Request Time
\end{minipage} & \begin{minipage}[b]{\linewidth}\raggedright
The First F1-Score Attack
\end{minipage} & \begin{minipage}[b]{\linewidth}\raggedright
The Average Attack Rate of The First Attacking Ip In The Future
\end{minipage} \\
\midrule
\endhead
1 & 3 & 2942 & 1755 & 3034 & 519 & 0.99 & 0.985 \\
2 & 3 & 2993 & 1845 & 3053 & 561 & 0.99 & 0.970 \\
3 & 3 & 3247 & 1548 & 2712 & 582 & 1 & 0.985 \\
Average & & 3061 & 1716 & 2933 & 554 & 0.99 & 0.980 \\
\bottomrule
\end{longtable}

Under a low attack ratio, the initial surge in a large number of
requests indicates the activation of the auto-scaling function to handle
high concurrent traffic. However, since the HPA's maxReplicas trigger
value is set to an attack ratio of 1\% over the last minute, after the
first trigger, the HPA maxReplicas value is set to 1. As a result, the
business pod is scaled down to 1, leading to a rapid decline in system
service capacity, and the number of 5XX errors continues to increase.

This situation persists until the second execution of the machine
learning script, which calculates that the overall attack ratio is still
above the threshold. Consequently, the service pod remains set to 1,
causing a large number of requests to fail, resulting in a significant
increase in 5XX status codes as well as 499 status codes. However, the
traffic from attacking IPs is correctly redirected to the Honeypot pod.

Table 5: Result of condition 4

\begin{longtable}[]{@{}
  >{\raggedright\arraybackslash}p{(\columnwidth - 14\tabcolsep) * \real{0.0886}}
  >{\raggedright\arraybackslash}p{(\columnwidth - 14\tabcolsep) * \real{0.0908}}
  >{\raggedright\arraybackslash}p{(\columnwidth - 14\tabcolsep) * \real{0.1579}}
  >{\raggedright\arraybackslash}p{(\columnwidth - 14\tabcolsep) * \real{0.1249}}
  >{\raggedright\arraybackslash}p{(\columnwidth - 14\tabcolsep) * \real{0.1735}}
  >{\raggedright\arraybackslash}p{(\columnwidth - 14\tabcolsep) * \real{0.1020}}
  >{\raggedright\arraybackslash}p{(\columnwidth - 14\tabcolsep) * \real{0.1102}}
  >{\raggedright\arraybackslash}p{(\columnwidth - 14\tabcolsep) * \real{0.1522}}@{}}
\toprule
\begin{minipage}[b]{\linewidth}\raggedright
Number
\end{minipage} & \begin{minipage}[b]{\linewidth}\raggedright
Condition
\end{minipage} & \begin{minipage}[b]{\linewidth}\raggedright
Nginx Attacks Received(Second layer pod)
\end{minipage} & \begin{minipage}[b]{\linewidth}\raggedright
5XX Number
\end{minipage} & \begin{minipage}[b]{\linewidth}\raggedright
Honeypot Attacks Received(Isolating pods)
\end{minipage} & \begin{minipage}[b]{\linewidth}\raggedright
Total Request Time
\end{minipage} & \begin{minipage}[b]{\linewidth}\raggedright
The First F1-Score Attack
\end{minipage} & \begin{minipage}[b]{\linewidth}\raggedright
The Average Attack Rate of The First Attacking Ip In The Future
\end{minipage} \\
\midrule
\endhead
1 & 4 & 2711 & 540 & 3192 & 362 & 0.93 & 0.911 \\
2 & 4 & 3400 & 662 & 2498 & 351 & 0.99 & 0.988 \\
3 & 4 & 2641 & 635 & 3189 & 362 & 0.99 & 0.979 \\
Average & & 2917 & 612 & 2960 & 358 & 0.97 & 0.959 \\
\bottomrule
\end{longtable}

Under a low attack ratio, the rapid increase in requests at the start
indicates that the auto-scaling function is triggered to handle the high
concurrent traffic. However, since the HPA's maxReplicas trigger value
is set to an attack ratio of 5\% over the last minute, the HPA's
maxReplicas value was not modified during the two script executions.

After the first script execution, all attack traffic was redirected to
the honeypot pods, resulting in no further increase in 5XX and 4XX
status codes. The overall service speed remained high, and the total
time required to complete all requests was relatively low.

\includegraphics[width=2.49236in,height=2.27986in]{image8.png}
\includegraphics[width=2.49236in,height=2.25417in]{image9.png}

Figure 9: The results of Condition 5 in Grafana Figure 10: The results
of Condition 6 in Grafana

Table 6: Result of condition 5

\begin{longtable}[]{@{}
  >{\raggedright\arraybackslash}p{(\columnwidth - 14\tabcolsep) * \real{0.0886}}
  >{\raggedright\arraybackslash}p{(\columnwidth - 14\tabcolsep) * \real{0.0908}}
  >{\raggedright\arraybackslash}p{(\columnwidth - 14\tabcolsep) * \real{0.1579}}
  >{\raggedright\arraybackslash}p{(\columnwidth - 14\tabcolsep) * \real{0.1249}}
  >{\raggedright\arraybackslash}p{(\columnwidth - 14\tabcolsep) * \real{0.1735}}
  >{\raggedright\arraybackslash}p{(\columnwidth - 14\tabcolsep) * \real{0.1020}}
  >{\raggedright\arraybackslash}p{(\columnwidth - 14\tabcolsep) * \real{0.1102}}
  >{\raggedright\arraybackslash}p{(\columnwidth - 14\tabcolsep) * \real{0.1522}}@{}}
\toprule
\begin{minipage}[b]{\linewidth}\raggedright
Number
\end{minipage} & \begin{minipage}[b]{\linewidth}\raggedright
Condition
\end{minipage} & \begin{minipage}[b]{\linewidth}\raggedright
Nginx Attacks Received(Second layer pod)
\end{minipage} & \begin{minipage}[b]{\linewidth}\raggedright
5XX Number
\end{minipage} & \begin{minipage}[b]{\linewidth}\raggedright
Honeypot Attacks Received(Isolating pods)
\end{minipage} & \begin{minipage}[b]{\linewidth}\raggedright
Total Request Time
\end{minipage} & \begin{minipage}[b]{\linewidth}\raggedright
The First F1-Score Attack
\end{minipage} & \begin{minipage}[b]{\linewidth}\raggedright
The Average Attack Rate of The First Attacking Ip In The Future
\end{minipage} \\
\midrule
\endhead
1 & 5 & 3030 & 1392 & 2984 & 429 & 1 & 1 \\
2 & 5 & 2685 & 1986 & 3133 & 422 & 0.95 & 0.958 \\
3 & 5 & 2898 & 1672 & 3243 & 423 & 0.99 & 0.969 \\
Average & & 2871 & 1689 & 3111 & 425 & 0.98 & 0.976 \\
\bottomrule
\end{longtable}

At the beginning of a large number of requests, there is a rapid
increase, indicating the triggering of the auto-scaling function to
handle high concurrent traffic. Since the HPA's maxReplicas trigger
value is set to an attack ratio of 1\% over the last minute, after the
first script execution, due to the 3\% attack ratio exceeding the 1\%
threshold, the HPA maxReplicas value was adjusted to 1. This caused an
increase in 5XX errors and 499 status codes due to the reduced service
capacity and high concurrency. During the second script execution, the
traffic from attacking IPs was redirected to honeypot pods, resulting in
no detected attacks in the last minute. Consequently, the HPA
maxReplicas value was adjusted to 5, fully restoring the system's
service capacity.

Table 7: Result of condition 6

\begin{longtable}[]{@{}
  >{\raggedright\arraybackslash}p{(\columnwidth - 14\tabcolsep) * \real{0.0886}}
  >{\raggedright\arraybackslash}p{(\columnwidth - 14\tabcolsep) * \real{0.0908}}
  >{\raggedright\arraybackslash}p{(\columnwidth - 14\tabcolsep) * \real{0.1579}}
  >{\raggedright\arraybackslash}p{(\columnwidth - 14\tabcolsep) * \real{0.1249}}
  >{\raggedright\arraybackslash}p{(\columnwidth - 14\tabcolsep) * \real{0.1735}}
  >{\raggedright\arraybackslash}p{(\columnwidth - 14\tabcolsep) * \real{0.1020}}
  >{\raggedright\arraybackslash}p{(\columnwidth - 14\tabcolsep) * \real{0.1102}}
  >{\raggedright\arraybackslash}p{(\columnwidth - 14\tabcolsep) * \real{0.1522}}@{}}
\toprule
\begin{minipage}[b]{\linewidth}\raggedright
Number
\end{minipage} & \begin{minipage}[b]{\linewidth}\raggedright
Condition
\end{minipage} & \begin{minipage}[b]{\linewidth}\raggedright
Nginx Attacks Received(Second layer pod)
\end{minipage} & \begin{minipage}[b]{\linewidth}\raggedright
5XX Number
\end{minipage} & \begin{minipage}[b]{\linewidth}\raggedright
Honeypot Attacks Received(Isolating pods)
\end{minipage} & \begin{minipage}[b]{\linewidth}\raggedright
Total Request Time
\end{minipage} & \begin{minipage}[b]{\linewidth}\raggedright
The First F1-Score Attack
\end{minipage} & \begin{minipage}[b]{\linewidth}\raggedright
The Average Attack Rate of The First Attacking Ip In The Future
\end{minipage} \\
\midrule
\endhead
1 & 6 & 3777 & 366 & 2168 & 339 & 1 & 0.99 \\
2 & 6 & 2938 & 441 & 3028 & 346 & 1 & 1 \\
3 & 6 & 2826 & 431 & 3105 & 348 & 1 & 1 \\
Average & & 3180 & 413 & 2767 & 344 & 1 & 1 \\
\bottomrule
\end{longtable}

Under a low attack ratio, the rapid increase in requests at the start
indicates the triggering of the auto-scaling function to handle high
concurrent traffic. However, since the HPA's maxReplicas trigger value
is set to an attack ratio of 5\% over the last minute, after the first
script execution, the 3\% attack ratio was below the 5\% threshold.
Consequently, after the first attack, all attack traffic was redirected
to the honeypot pods, resulting in no new 5XX or 4XX errors.

\hypertarget{experimental-results}{%
\section{Experimental Results}\label{experimental-results}}

\includegraphics[width=5.09167in,height=3.39444in]{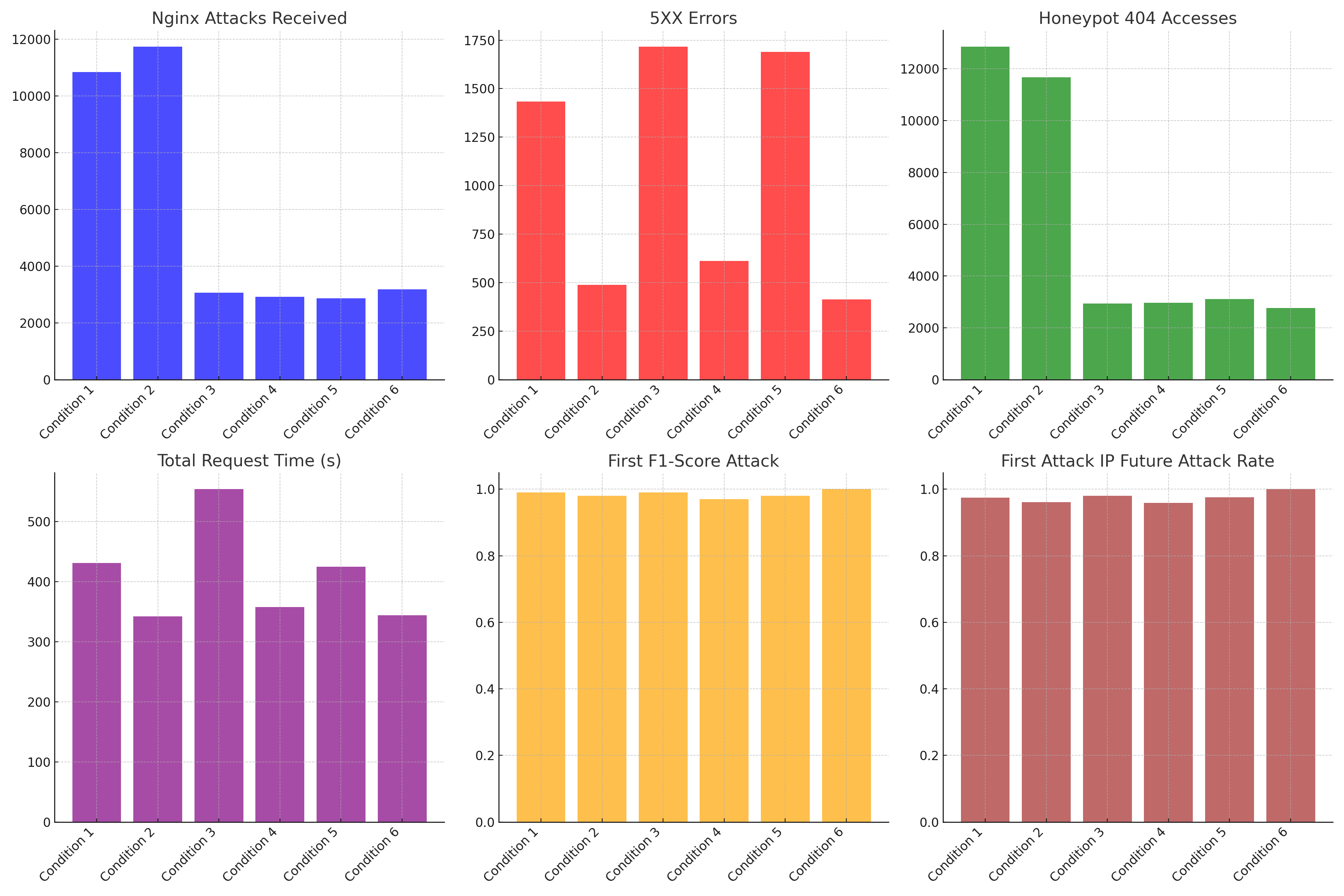}

Figure 11: Experimental results bar graph

From this Figure 11, it can be observed that the number of attacks for
conditions 1 and 2 is relatively high because their designed attack
ratio is 12\%. The attack volume for other conditions is lower. The 5XX
ratio shows that conditions 1, 3, and 5 have higher values. This is
because these three conditions reached the HPA's maxReplicas trigger
value, causing the pod to be reduced to a minimum of 1. The honeypot
also received 404 accesses similarly. After executing the machine
learning script, the number of 404 attacks on the honeypot was similar
to the attack situation on the actual service pod.

In terms of overall service time, conditions 2, 4, and 6 did not trigger
the HPA's maxReplicas adjustment, resulting in the shortest service
times. In other conditions, the HPA's maxReplicas adjustment was
triggered, leading to longer overall request times, especially for
condition 3. Since the second execution of the machine learning script
did not meet the conditions to lift the attack status, both its service
time and 5XX errors were the longest. The attack simulation using
scripts has certain limitations, but the F1 indicator and future
predictions are relatively high. This suggests that using the random
forest algorithm is relatively suitable for directory scanning
attacks[22].

Due to the small size of the experimental data, there is an increased
possibility of overfitting. However, the risk is reduced by the limited
selection of features. The main focus of this paper is on how to
integrate machine learning models with HPA components to achieve attack
detection, prediction, and isolation, and dynamically adjust HPA
parameters. The goal is for this framework to be applicable to a wider
range of scenarios in the future. The issue of overfitting will be
further explored and optimized in future research, potentially using
techniques such as regularization or data augmentation to improve the
machine learning models. The current experimental results demonstrate
the framework's validity and its potential for future scalability.

Therefore, the conclusion drawn from this experiment is that the
designed framework can effectively adjust the number of pods under high
load conditions and detect scanning attacks while adjusting HPA
parameters to prevent the waste of hardware resources due to the
continuous increase in pod numbers caused by attacks. When an attack is
detected, this framework can effectively redirect all traffic from the
attacking IP to the honeypot pod, providing isolation and allowing for
further behavioral analysis in the future. However, this experiment
demonstrates that the threshold settings for HPA's maxReplicas
adjustment are crucial. Too high or too low thresholds fail to achieve
the expected results, thus requiring reasonable settings based on actual
attack conditions.

\hypertarget{conclusion}{%
\section{CONCLUSION}\label{conclusion}}

This study proposes the use of Random Forest machine learning analysis
for attack detection, redirecting the corresponding attacks into
honeypot pods, and dynamically adjusting HPA parameters based on the
attack situation. This approach achieves effective automatic scaling of
Nginx services in Kubernetes while effectively preventing and managing
attacks. Using 5XX status codes as custom metrics aligns more closely
with real service scenarios. This study experimentally integrates
machine learning with Kubernetes HPA, and the framework can be adapted
to use other machine learning models in the future for effective
detection and prediction of different types of attacks.

\end{document}